\documentclass[11pt]{article}
\usepackage{graphicx,graphics,times}
\usepackage{bm}
\usepackage{longtable}

\def\be{\begin{equation}}
\def\ee{\end{equation}}
\def\ba{\begin{eqnarray}}
\def\ea{\end{eqnarray}}
\def\la{\langle}
\def\ra{\rangle}
\def\a{\alpha}
\def\b{\beta}

\begin{document}

\begin{titlepage}
\vspace{4cm}
\begin{center}{\Large \bf How the quality of a received EPR pair depends on the distances from an EPR source?}\\
\vspace{1cm}Abolfazl Bayat\footnote{email: bayat@mehr.sharif.edu},
\hspace{0.5cm} Vahid Karimipour \footnote{Corresponding author:vahid@.sharif.edu}\\
\hspace{0.5cm} Iman Marvian \footnote{iman@mehr.sharif.edu}\\
\vspace{1cm} Department of Physics, Sharif University of Technology,\\
P.O. Box 11365-9161,\\ Tehran, Iran
\end{center}
\vskip 3cm
\begin{abstract}
Let an EPR source which generates maximally entangled pairs be
located so that it has distances $L_1$ and $L_2$ to two users. After
taking into account various effects like loss of photons,
deficiencies in the source and detectors, an entangled pair
traveling through the channel may loose its perfect correlation due
to errors in the channel. How the entanglement of the received pair
depends on the above distances and the local properties of the
channels used for this transmission? What is the best location of
the source if we want to achieve the highest fidelity? What is the
threshold distance beyond which the entanglement of the pair
vanishes and becomes useless for using in teleportation. We discuss
these problems for the Pauli channel which simulates the effect of
optical fibers and possibly the atmosphere on the
polarization-entangled photons.
\end{abstract}

\vskip 2cm PACS Numbers: 03.65.Ud
\end{titlepage}

\date{\today}

\section{Introduction} Almost any protocol of quantum information processing
requires an entangled pair of particles. Traditionally these pairs
have been known as EPR pairs after Einstein, Podolsky and Rosen
\cite{EPR} or Bell states after John Bell \cite{bell}. It is usually
envisaged that there is an EPR source generating EPR states which
distributes such pairs to various users or communicating parties
\cite{ekert,bennett,dense,algorithm}. In recent years a great deal
of experimental effort has been devoted to production and
transmission of entangled photons \cite{tapster, weihs, jennewein,
wein, villoresi, vienna1, vienna2, marcikic, buttler, tittel,
aspelmeyer}. Although these efforts have initially been fueled by
the need to test Bell inequality and its generalizations, in more
recent years they are mostly motivated by the perspective of
possibility of long distance quantum communication and actual
implementation of some quantum information protocols.\\
This interest is so strong that some researchers are even
considering to go beyond Earth-bound laboratories and use satellites
as transmitter, receiver or relay terminals \cite{aspelmeyer, villoresi} for quantum communication. \\

Over the years it has become possible to distribute entangled
photons through increasingly long distances. Some of the distances
reported are 360 meters \cite{jennewein}, 400 meters \cite{weihs},
$1.45$ km \cite{vienna2}, and $4$ km \cite{tapster} for polarization
entangled photons through optical fibers, $10$ km \cite{tittel} and
$50$ km \cite{marcikic} for time-bin entangled photons through
optical fibers and $7.8$ km for polarization entangled photons
through free space \cite{ vienna1}.\\

In all these experiments various loss effects should be taken into
account. The first important loss occurs when one or both photons of
the pair are absorbed or scattered in the channel and do not reach
the detectors. This type of loss has an exponential dependence on
distance with a typical attenuation of a few dB per km depending on
the wavelength, (e.g. 3.2 dB/km for optical wavelengths
\cite{vienna2}, 0.35 dB/km for telecommunication wavelengths
\cite{marcikic}). Those truly entangled photons which reach the
detectors can be detected by precise coincidence counting in a
nanosecond window. The coincidence peak is nearly noise free with a
signal to noise ratio exceeding 100 \cite{weihs}. Detection is
usually done by silicon-avalanche photo diodes (SAPD) which have
dark counts of a few hundred per second, much lower than 10000-15000
counts in the experiment of Weihs et al in 1998 \cite{weihs} or
60000  count rates in more recent experiments of the Vienna group
\cite{vienna1}. One should of course take into account these dark
and accidental coincidence counts and subtract them from all the
measured coincidence
counts to find the actual number of entangled photons received. \\

Trying to increase the distance over which entangled photons are
distributed is a fairly difficult task. The basic problem is the
increased photon loss in the channel which lowers the signal to
noise ratio at the detectors. Trying to compensate for the photon
loss in the channel, one can raise the rate of photon production in
the parametric down conversion process, but this makes the source to
deviate more from an ideal source of EPR pairs. In fact the output
state of parametric down conversion is always an (ideal) Bell state
mixed with other no-photon or multi-photon states and a stronger
source raises the probability of these unwanted states. Therefore
one has to make a compromise between conflicting factors. A
theoretical analysis of the situation is therefore necessary to find
the actual limitations on quantum
communication tasks \cite{waks}.\\

In \cite{waks} a detailed study of quantum key distribution has been
performed taking into account real experimental situations. In that
work the main effects which are taken into account are the
deficiency of the parametric down conversion mechanism which
produces a Bell state only in a coherent superposition with other
un-wanted multi-photon states, deficiencies in the detectors which
produce dark counts and false coincidence counts and finally photon
loss in the channel which is considered to be exponential in terms
of the length of the channel and is modeled by a beam splitter. In
this way it has been shown that security of the BBM92 protocol
\cite{bbm92} for quantum key distribution can be attained to
distances up to 170 km with the assistance of entanglement swapping.
Moreover it has been shown that the loss factors are minimized if
the EPR source is situated midway between
the two parties involved in this protocol.\\

Overall it seems that with current technology the earth-bound fibre
and free-space quantum communication cannot surpass on the order of
100km [16]. \\

After taking all the above losses into account, we are still faced
with the problem that such pairs may loose their perfect correlation
or anti-correlation in passing through the channel from the source
to the detectors.

Usually polarization drifts or partial rotations of photon
polarizations from the source to the fibers are corrected by passing
them through compensators and a software which calculates these
drifts is used to adjust the phase of the singlet and to compensate
the in-crystal birefringence \cite{vienna1, vienna2, weihs}. However
this compensation is only done at the beginning of the channel and
not the whole length of the channel. Even this, is not always easy.
For example in a recent experiment of the Vienna group in which
entangled photons are produced and sent to a distance of 7.8 km
through air, it is argued that the final states may have been more
entangled than indicated by their measured violation of Bell
inequality due to the partial rotation of polarization states
\cite{vienna1}.\\

Unfortunately a characterization of quantum channels in terms of
polarization drift does not exists and one can only infer the error
rates produced in such channels by subtracting known error rates
from the measured loss of visibility. For example in the experiment
of \cite{ vienna2} with a distance of $7.8$ km between the source
and the receiver, after entangled photons have been extracted by
precise coincidence counting, the average loss of visibility has
been around 8 percent, 2.5 percent of which has been accounted for
by imperfection of detectors, 1.2 percent by imperfection of the
source and the rest has been attributed to the quantum channel. In
another experiment \cite{weihs} with a distance of 400 m, it has
been found that
polarization drift has been less than 1 percent. \\
This is all for fibers, and for free air it is known that the
atmosphere is almost non-birefringent. \\

However if we are going to break the barrier of 100 km and do long
distance quantum communication by advancing our technology,
certainly the errors in the channel no matter how small they are per
length, should be calculated and taken into account. In this article
we want to see how the entanglement of an EPR pair distributed by a
source, diminishes when the two qubits (e.g. polarization states of
photons) travel through distances $L_1$ and $L_2$ in a Pauli channel
to reach the users. Such a channel has been shown to be suitable for
modeling thermally fluctuating birefringence of
single mode fibers carrying the polarization states of photons \cite{depol}. \\

The main quantities that we calculate are the fidelity of the
received EPR pair with the initially maximally entangled pair and
also the concurrence of the former which turn out to be functions of
$L_1+L_2$. We will find the threshold for $L_1+L_2$, beyond which an
EPR pair looses all its entanglement, as a function of the error
parameters of the channel. We will find that if in the Pauli channel
only one bit flip error occurs, then the concurrence of the received
pair never vanishes except for infinitely long channels. However if
two or three bit-flip errors occur, then there is always a threshold
length beyond which entanglement of the received pair vanishes. From
current experimental data we estimate that for the depolarization
channel, as a model for optical fibers, this threshold distance is
more than 34 kilometers.

The rest of this paper is structured as follows:  We first determine
in section \ref {sec1} the dependence of total error parameters of
the Pauli channel as a function of its length and the value of its
local error parameters.  Then in section \ref{sec2} we send a
maximally entangled pair through the channel and calculate the
fidelity of the final received pair, with the initial pair.

\section{Characterization of the Pauli channel error rates in terms of its length}\label{sec1}
 A quantum channel is specified by the types of error
operators it introduces on arbitrary states which are transmitted
through the channel. In practice we can characterize a short
segment of a channel by suitable quantum measurements in the
laboratory. It is then possible to infer the characteristics of an
arbitrary length of the channel by combining the quantum
operations which pertain to each segment. Suppose a segment of the
channel is specified by a quantum operation given by the Kraus
decomposition
\begin{equation}\label{kraus}
    \rho_1  \equiv E(\rho)=  \sum_{k} E_k \rho E_{k}^{\dagger}.
\end{equation}
Then a sequence of $N$ segments of such a channel is specified by
the quantum operation
\begin{equation}\label{kraus2}
    \rho_N \equiv E^{N}(\rho) = E(E(\cdots E(\rho))).
\end{equation}
If we allow for a large enough set of operators, a large enough
parameter space, we can say that the set of quantum operations is
closed under this concatenation. The problem then reduces to finding
the flow of parameters under concatenation. By going to the limit of
an infinite number of infinitesimal segments, we can obtain the
parameters as functions of the length of the channel (or the time
duration of the channel for those cases where our channel only
stores data).

An important example of a quantum channel acting on a qubit is the
Pauli channel which is specified by the following quantum operation
\begin{equation}\label{pauli1}
   \rho_1:= E(\rho)=p_0 \rho + p_1 \sigma_x \rho \sigma_x + p_2 \sigma_y \rho \sigma_y + p_3 \sigma_z \rho
    \sigma_z,
\end{equation}
where $p_i, \ i=0, \cdots ,3$ are respectively the probabilities
of no error, bit-flip, bit-phase-flip and phase-flip errors
\cite{Nielsen,preskill} and $p_0 + p_1+p_2+p_3=1$.

Recently it has been shown that this channel can model the effects
of some realistic noise on qubits, like the effect of randomly
fluctuating magnetic fields on electron spin or thermally induced
birefringence of polarization states of photons traveling through
optical fibers \cite{depol}.\\

Using the properties of Pauli matrices, it is easily verified that
the concatenation of two Pauli channels is again a Pauli channel.
Therefore we can write
\begin{equation}\label{pauliN}
    E^N(\rho)=p_0^{(N)} \rho + p_1^{(N)} \sigma_x \rho \sigma_x + p_2^{(N)} \sigma_y \rho \sigma_y +
    p_3^{(N)} \sigma_z \rho
    \sigma_z.
\end{equation}
Using the relation $E^{N+1}(\rho)=E(E^{N}(\rho))$ and the properties
of the Pauli matrices, a recursion relation between the error
parameters is obtained. Written in matrix from it reads
\begin{equation}\label{recursion}
\left(
\begin{array}{c}
  p_0^{(N+1)} \\
  p_1^{(N+1)} \\
  p_2^{(N+1)} \\
  p_3^{(N+1)} \\
\end{array}
\right)= \left(
\begin{array}{cccc}
  p_0 & p_1 & p_2 & p_3 \\
  p_1 & p_0 & p_3 & p_2 \\
  p_2 & p_3 & p_0 & p_1 \\
  p_3 & p_2 & p_1 & p_0 \\
\end{array}
\right)\left(
\begin{array}{c}
   p_0^{(N)} \\
  p_1^{(N)} \\
  p_2^{(N)} \\
  p_3^{(N)} \\

\end{array}
\right).
\end{equation}

This relation can be solved by diagonalizing the matrix and using
the boundary conditions $p_{i}^{(0)}=\delta_{i,0},
 i=0,1,2,3$. The final result can be written compactly using a matrix notation, i.e
\begin{equation}\label{lambda}
\left(\begin{array}{c}
       p_0^{(N)} \\
       p_1^{(N)} \\
       p_2^{(N)} \\
       p_3^{(N)} \\
      \end{array}
\right) = \frac{1}{4}\left(
\begin{array}{cccc}
 1 & \ \ 1 & \ \ 1 & \ \ 1 \\
  1 & \ \ 1 & -1 & -1 \\
  1 & -1 & \ \ 1 & -1 \\
  1 & -1 & -1 & \ \ 1 \\
\end{array}
\right) \left(\begin{array}{c}
        1 \\
        \lambda_1 \\
        \lambda_2 \\
        \lambda_3 \\
      \end{array}\right),
\end{equation}
where
\begin{eqnarray}\label{lambdas}
  \lambda_1 &=&(1-2p_2-2p_3)^N  \cr
  \lambda_2 &=& (1-2p_1-2p_3)^N \cr
  \lambda_3 &=& (1-2p_1-2p_2)^N .
\end{eqnarray}

This shows how the parameters of a channel made of $N$ consecutive
segments are related to the parameters of short segments which are
usually easier to characterize in practice. If one defines the
parameters $\mu_i:= \frac{p_i}{l}$ as error per length for very
short segments, then the parameters for a channel of length $L$
take the following form
\begin{equation}\label{lambdas3}
  \lambda_1= e^{-2(\mu_2+\mu_3)L},\ \ \ \lambda_2 = e^{-2(\mu_1+\mu_3)L}, \ \ \ \ \lambda_3 = e^{-2(\mu_1+\mu_2)L}  .
\end{equation}
Inserting the above values of $\lambda_i$ in (\ref{lambda}) gives
the error parameters of the channel in terms of its length.
\begin{equation}\label{lambdas2}
\left(\begin{array}{c}
       p_0(L) \\
       p_1(L) \\
       p_2(L) \\
       p_3(L) \\
      \end{array}
\right) = \frac{1}{4}\left(
\begin{array}{cccc}
 1 & \ \ 1 & \ \ 1 & \ \ 1 \\
  1 & \ \ 1 & -1 & -1 \\
  1 & -1 & \ \ 1 & -1 \\
  1 & -1 & -1 & \ \ 1 \\
\end{array}
\right) \left(\begin{array}{c}
        1 \\
        e^{-2(\mu_2+\mu_3)L} \\
        e^{-2(\mu_1+\mu_3)L} \\
        e^{-2(\mu_1+\mu_2)L} \\
      \end{array}\right).
\end{equation}

As a special case we will have the following result for flip
channels, where $\rho_{flip}(L)$ denotes the density matrix at
 the output of a channel of length $L$ and the index $i$ takes the values $1, 2$ and $3$ for the bit-flip,
 the bit-phase flip and the phase-flip channel respectively:
\begin{equation}\label{rhoLbit1}
    \rho_{flip}(L)= \frac{1}{2}(1+e^{-2\mu_i L})\rho + \frac{1}{2}(1-e^{-2\mu_i
    L})\sigma_i \rho \sigma_i.
\end{equation}
We can see that for very long channels where $L\rightarrow \infty
$, and we will  have
$\rho_{flip}(\infty)=\frac{1}{2}(\rho+\sigma_i \rho \sigma_i)$.
Thus the probability of flipping tends to $\frac{1}{2}$ as
expected for the worst case of a flip-channel. \\
Another special case is the depolarizing channel which is defined
by the quantum operation \cite{Nielsen,preskill}
\begin{equation}\label{depo}
    E(\rho)= p\frac{I}{2} + (1-p)\rho,
\end{equation}
or equivalently by equation (\ref{pauli1} ) with parameters
$p_0=1-\frac{3p}{4}, p_1=p_2=p_3=\frac{p}{4}$.  Using
(\ref{lambdas2} ) with $\mu:=\mu_1=\mu_2=\mu_3 $ we find
${P_0,}_{{\rm depol}}(L) = \frac{1}{4}(1+3e^{-4\mu L})$ and
${P_i,}_{{\rm depol}}(L)= \frac{1}{4}(1-e^{-4\mu L}), \ \  i=1, 2,
3$. Using the well known identity $\sum_{i} \sigma_i \rho \sigma_i =
2I - \rho$, we can rewrite this as
\begin{equation}\label{rhoLbit2}
    \rho_{depol}(L)= e^{-4\mu L} \rho + \frac{1}{2}(1-e^{-4\mu
    L})I.
\end{equation}
As $L\rightarrow \infty $ the probability of error tends to $1$
and we will have $\rho_{depol}(\infty)=\frac{I}{2}$ which is a
completely mixed state carrying no information of the original
state.

\section{Transmission of EPR pairs through Pauli channels}\label{sec2}
 We now consider two users with distances
$L_1$ and $L_2$ to an EPR source (figure(I)). The source prepares
a maximally entangled pair and sends each qubit of the pair to a
user. We want to calculate the efficiency of this process,
measured by the concurrence of the received pair as a function of
distances $L_1$ and $L_2$ and the error parameters of the channel,
which are assumed to be of the same type for both routes.
\begin{figure}\label{Alice}
\centering
    \includegraphics[width=8cm,height=4cm,angle=0]{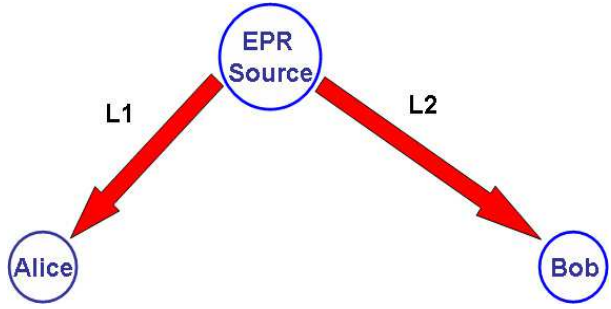}
    \caption{"(Color online)" Transmission of EPR pairs from a source to Alice and Bob
     trough channels with different lengths $L_1$, $L_2$.   }
\end{figure}

Suppose that the source sends a maximally entangled state
$\rho=|\psi^+\ra\la \psi^+|$ into the channel where $|\psi^+\ra =
\frac{1}{\sqrt{2}}(|0,0\ra + |1,1\ra)$. The parameters of the two
channels are denoted respectively by $r_i:=p_i(L_1)$ and $
s_i=p_i(L_2)$ derived from equation (\ref{lambdas2}).
 After transmitting the state through the Pauli channel, this state is transformed to
\begin{equation}\label{Epauli}
    \rho'= r_0s_0 \rho + r_0\sum_{k=1}^3 s_k (I\otimes \sigma_k)\rho(I\otimes
    \sigma_k)  +
    s_0\sum_{k=1}^3 r_k (\sigma_k\otimes I)\rho(\sigma_k\otimes
    I) +\sum_{k,l=1}^3 r_ks_l (\sigma_k\otimes \sigma_l)\rho(\sigma_k\otimes
    \sigma_l).
\end{equation}
Evaluation of the right hand side is facilitated by noting that
$\rho=|\psi^+\ra\la \psi^+|$ can be written as
$\rho=\frac{1}{2}(S-\sigma_y\otimes \sigma_y)$, where $S$ is the
swap operator $S|\a,\b\ra = |\b, \a\ra$. One then uses the following
easily verified identities
\begin{equation}\label{ide}
    \sigma_k \sigma_y \sigma_k = (-1)^k \sigma_y, \ \ \ \ S(A\otimes
    B)=(B\otimes A)S,
\end{equation}
and $\sum_{k=1}^3 \sigma_k\otimes \sigma_k = 2S-I$. \\ A rather
lengthy but straightforward calculation will determine the output
density matrix. It is given by
\begin{equation}\label{Epauliresult}
    \rho'= a |\psi^+\ra \la \psi^+| + b|\psi^-\ra \la \psi^-|+ c|\phi^+\ra \la \phi^+|+d|\phi^-\ra \la
    \phi^-|,
\end{equation}
where
\begin{eqnarray}
 \nonumber
  |\psi^{\pm}\ra &=& \frac{1}{\sqrt{2}}(|0,0\ra \pm |1,1\ra),  \\
  |\phi^{\pm}\ra &=& \frac{1}{\sqrt{2}}(|0,1\ra \pm |1,0\ra),
\end{eqnarray}
are the Bell states and
\begin{eqnarray}\label{a}
  a &=& r_0s_0 + r_1s_1 + r_2s_2 + r_3s_3 \cr
  b &=& r_0s_3 + r_1s_2 + r_2s_1 + r_3s_0 \cr
  c &=& r_0s_1 + r_1s_0 + r_2s_3 + r_3s_2 \cr
  d &=& r_0s_2 + r_1s_3 + r_2s_0 + r_3s_1.
\end{eqnarray}
These parameters are in fact the fidelities of the output state with
the Bell states $|\psi^+\ra,\  |\psi^-\ra,\  |\phi^+\ra\ $ and
$|\phi^-\ra$ respectively. \\
 Using the
relations (\ref{lambdas2}), (\ref{a} ) take the following form
\begin{eqnarray}\label{coeflen}
  a &=& \frac{1}{4}\{1+e^{-2(\mu_1+\mu_2)L}+e^{-2(\mu_1+\mu_3)L}+e^{-2(\mu_2+\mu_3)L} \}  \cr
  b &=& \frac{1}{4}\{1+e^{-2(\mu_1+\mu_2)L}-e^{-2(\mu_1+\mu_3)L}-e^{-2(\mu_2+\mu_3)L} \}
  \cr
  c &=& \frac{1}{4}\{1-e^{-2(\mu_1+\mu_2)L}-e^{-2(\mu_1+\mu_3)L}+e^{-2(\mu_2+\mu_3)L} \}
  \cr
  d &=& \frac{1}{4}\{1-e^{-2(\mu_1+\mu_2)L}+e^{-2(\mu_1+\mu_3)L}-e^{-2(\mu_2+\mu_3)L}
  \},
\end{eqnarray}
where $L:=L_1+L_2$. It is obvious that $a>b,c,d$ so the fidelity
of transmitted pair with $|\psi^+ \ra$ is greater than the other
Bell states.\\
We are interested in the concurrence of the state $\rho'$
\cite{wootters}, since it is a measure of the degree of mutual
entanglement of the received pair of qubits and hence a measure of
the success of any quantum protocol like teleportation which may use
this pair. \\
Using the result of \cite{wootters}, the concurrence $C(\rho')$ can
be calculated in a straightforward way. It is given by
\begin{eqnarray}\label{conc} C(\rho') = max(0,2
\alpha_{max}-1),
\end{eqnarray}
where  $\alpha_{max}:={\rm max}(a,b,c,d)$ . Note that a nonzero
concurrence implies that $\alpha_{max}> \frac{1}{2}$ which in turn
implies that the fidelity of the state $\rho'$ with one of the Bell
states is greater than $\frac{1}{2}$. Under these conditions one can
use the state $\rho'$ to achieve a fidelity of teleportation
exceeding the one obtained in the best classical protocols
\cite{horo,classictele}. Using (\ref{coeflen}) and (\ref{conc}) we
obtain
\begin{equation}\label{conclen}
C_{\rm
Pauli}(\rho')=max(0,\frac{1}{2}\{e^{-2(\mu_1+\mu_2)L}+e^{-2(\mu_1+\mu_3)L}+e^{-2(\mu_2+\mu_3)L}-1\})
\end{equation}
Thus the fidelity and the concurrence depend only on the sum of
the distances and not on the individual distances.  Specially if
the EPR source is collinear with the users and situated between
them, then the location of the source is immaterial to the
efficiency.\\

\subsection{Single Bit-flip channels}

As a special case we study the bit-flip channel for which
$\mu_2=\mu_3=0$. Equations (\ref{Epauliresult}) and (\ref{coeflen})
show that
\begin{equation}\label{new}
    a=\frac{1}{2}(1+e^{-2\mu L}), \ \ \ b=0, \ \ \ c=\frac{1}{2}(1-e^{-2\mu
    L}), \ \ \ d=0.
\end{equation} Using (\ref{Epauliresult}, \ref{a}) and
(\ref{conclen}) we find the fidelity of the output state with the
initial Bell state $|\psi^+\ra$ and its concurrence to be
\begin{eqnarray}\label{conclen1}
\la \psi^+|\rho'|\psi^+\ra &=& \frac{1}{2}(1+2^{-2\mu_L})\cr C_{\rm
bit-flip}(\rho')&=& e^{-2\mu_1L}.
\end{eqnarray}
This shows that for the bit flip channel, no matter how long the
channel is, the received state can always be used for teleporation
or some other quantum protocol, possibly after some distillation to
increase the efficiency. This result is also valid for the other
single flip-channels. Note that in this case the output density
matrix is a mixture of only two Bell states and this mixture is
50-50 only when the length of the channel goes to infinity. This is
in agreement with a result of Horodecki's \cite{hor} which state
that a mixture of two Bell states is always entangled , except when
the mixture is
50-50.\\
\subsection{Double bit-flip channels} If the channel allows for more
than one type of flip error, then there is always a threshold
distance beyond which the received state is useless. To see this
consider the case where $\mu_3=0$ and $\mu_1=\mu_2=\mu$. In this
case we find from (\ref{conclen}) that
\begin{equation}\label{new2}
    a=\frac{1}{4}(1+e^{-2\mu L})^2, \ \ \ b=\frac{1}{4}(1-e^{-2\mu L})^2, \ \ \ c=d=\frac{1}{4}(1-e^{-4\mu
    L}).
\end{equation}

From this we obtain
\begin{eqnarray}\label{conclen2}
\la \psi^+|\rho'|\psi^+\ra &=& \frac{1}{4}(1+e^{-2\mu L})^2\cr
 C_{\rm
double-flip}(\rho')&=&\frac{1}{2}max(0,e^{-4\mu L}+2e^{-2\mu L}-1),
\end{eqnarray}
which implies that beyond a threshold length
\begin{equation}\label{length2}
    L^{{\rm th}}_{double-flip} := \frac{1}{2\mu}\ln(\frac{1}{\sqrt{2}-1})
\end{equation}
the concurrence vanishes.\\
\subsection{Depolarization Channel} For the depolarizing channel
where $\mu_1=\mu_2=\mu_3=\mu$ we find from (\ref{conclen}) that
\begin{equation}\label{new3}
    a=\frac{1}{4}(1+3e^{-4\mu L}), \ \ \ \  b=c=d=\frac{1}{4}(1-e^{-4\mu
    L}).
\end{equation}
From this we obtain
\begin{eqnarray}\label{conclen3}
\la \psi^+|\rho'|\psi^+\ra &=& \frac{1}{4}(1+3e^{-4\mu L})\cr C_{\rm
depol}(\rho')&=&\frac{1}{2}max(0,3e^{-4\mu L}-1),
\end{eqnarray}
which implies that beyond a threshold length
\begin{equation}\label{length3}
    L^{{\rm th}}_{depol} := \frac{\ln 3}{4\mu}
\end{equation}
the concurrence vanishes, rendering the transmission useless. In
figure(II), $C_{depol}(\rho')$ is plotted in terms of $L$ for two
different values of $\mu$.
\begin{figure}\label{depol}
\centering
    \includegraphics[width=12cm,height=8cm,angle=0]{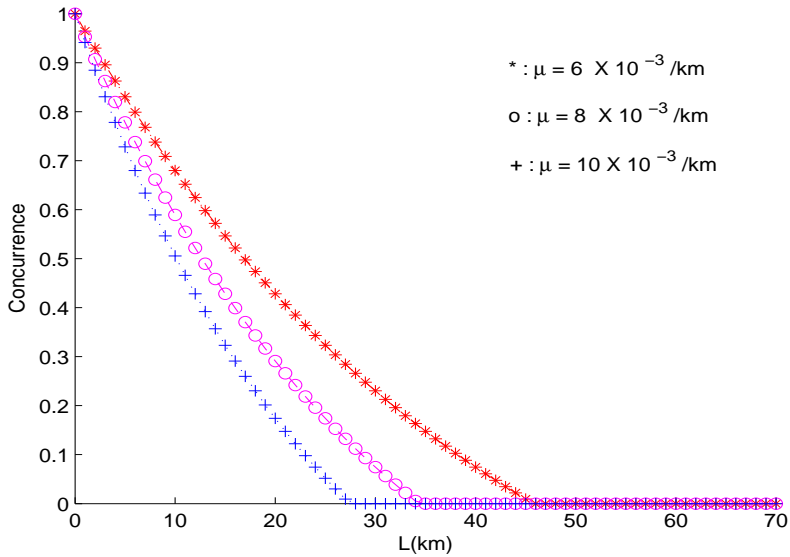}
    \caption{"(Color online)"  Concurrence versus total length $L=L_1+L_2$ in depolarizing channel
      for two different values of $\mu$ .}
\end{figure}

\section{Relation with experiments}
Assuming that a depolarization channel can model the polarization
drift of photons in an optical fiber, we can make a prediction as to
how long two photons can travel through this media before their
correlations are totally lost. In the experiment of Weihs et al
\cite{weihs}, it is reported that the polarization drift has been
less than 1 percent after the photons have traveled a distance of
400 meters. Inserting this value in equation (\ref{conclen3}) one
deduces a value of $\mu \approx 8 \times 10^{-3}/km$. In a more
recent experiment \cite{vienna1}, from an average QBER (qubit error
rate) of 8 percent, 2.5 percent is accounted for by imperfection of
detectors, 1.2 percent by imperfection of photon sources and the
rest 4.3 percent by the errors in the quantum channel. Inserting
this in equation (\ref{conclen3}) for $L=1.45 km$ \cite{vienna2},
gives a value of $\mu \approx 10 \times 10^{-3}/km.$ If we take a
tentative value of $\mu \sim 8 \times 10^{-3}/km $ for transmission
of polarization-entangled photons through optical fibers, then from
equation (\ref{length3}) we find a threshold distance of about $34
$km, beyond which the concurrence of the final EPR pair drops below
zero.\\ This value is much longer than the current distances over
which polarization-entangled photons have been distributed along
fibers, and lower than the distance limit of about 60 km that
time-bin entangled photons can be distributed \cite{marcikic}.
However it is less than the limit of 170 km found in \cite{waks}
which has been found mainly on the basis of a compromise between
increasing the photon production (to overcome absorbtion) and
increasing the efficiency of the EPR source (the parametric down
conversion process). At present experimental data can not determine
which of the above bounds is more stringent. Moreover if we note
that part of the errors in the cited experiments above is due to
polarization misalignments \cite{vienna2}, then the estimated value
of $\mu$ will become smaller leading to threshold distances longer
than 34 kilometers. 

\section{Discussion}
Given a transmission line of energy, i.e. electric power, optical
signals,  it is always possible to determine the loss of energy in
terms of the local properties of the line and the distance from the
generator. We have asked this same question for a class of quantum
channel, namely the Pauli channel and have derived expressions of
error parameters in terms of the local error densities (probability
of error per length) and the length of the channel. We have then
considered an EPR source, which is to send a maximally entangled
pair to two users which have distances $L_1$ and $L_2$ to the
source. We have calculated the entanglement of the received pair and
its fidelity with the original pair as a function of these
distances. For the Pauli channel the concurrence of the final
received pair depends only on $L_1+L_2$, the sum of the two
distances. In the special cases where the source is collinear with
the users and is located between the users, our results show that
for the Pauli channel, the efficiency is independent of the location
of the source, although to minimize other losses the best location
of the source turns out to be midway between the source and the
receivers \cite{waks}. By using some of the current experimental
data we have determined a threshold distance beyond which the
correlation of the initial EPR pair drops to zero. This distance
certainly is greater than 34 kilometers. In order to find more
precise values of threshold distances one should have a
characterization of local error parameters of optical fibers.

\section{Acknowledgement}
We thank A. Vaziri and A. Zeilinger for very helpful email
correspondences. We also thank M. Asoudeh , D. Lashkari, N. Majd ,
L. Memarzadeh, and A. Sheikhan, for valuable discussions. Finally we
wish to thank the anonymous referee for his very helpful comments
and criticism.


\begin{thebibliography}{}
\bibitem{EPR} A. Einstein, B. Podolsky, and N. Rosen, Phys. Rev.
, \textbf{47(10)}, 777 (1935).

\bibitem{bell}
 J. S. Bell, {\it Speakable and Unspeakable in Quantum Mechanics} (Cambridge University Press, Cambridge, 1987).

\bibitem{ekert} D. Bouwmeester, A. Ekert, A. Zeilinger, The Physics of Quantum
Information (Springer, Berlin, 2000).

\bibitem{bennett} C. H. Bennett, G. Brassard, C. Crépeau, R. Jozsa, A. Peres, and W. K. Wootters
Phys. Rev. Lett. \textbf{70}, 1895-1899 (1993)

\bibitem{dense} Bennett, C. H. and Wiesner, Phys. Rev.
Lett. \textbf{69}, 2881 (1992).

\bibitem{algorithm} A. Ekert, Phys. Rev. Lett. \textbf{67}, 661 (1991).

\bibitem{tapster} P. R. Tapster, J. G. Rarity and P. C. M. Owens,
Phys. Rev. Lett. {\bf 73}, 1923 (1994).

\bibitem{weihs} G. Weihs, T. Jennewein, C. Simon,
H. Weinfurter, A. Zeilinger, Phys.Rev.Lett. \textbf{81} (1998)
5039-5043 ,quant-ph/9810080.

\bibitem{jennewein} T. Jennewein, C. Simon, G.W eihs, H. WeinfurterD, A. Zeilinger,
 Phys. Rev. Lett. \textbf{84}, 4729 (2000),quant-ph/9912117.

\bibitem{wein} A.M. Pupasov, B.F. Samsonov, SIGMA, Vol. 1 (2005), Paper 020,
  quant-ph/0101074.

\bibitem{villoresi} Villoresi \emph{et. al.}, SPIE proceedings Quantum
Communications and Quantum Imaging II conference in Denver, July
2004, quant-ph/0408067.

\bibitem{vienna1} K.J.Resch , \emph{et.al} , Opt. Express 13, 202-209
(2005), quant-ph/ 0501008 .

\bibitem{vienna2} A. Poppe, \emph{et.al}
,Opt. Express 12, 3865-3871 (2004) ,quant-ph/0404115.

\bibitem{marcikic} I. Marcikic et al. Phys. Rev. Lett. {\bf 93}
180502 (2004).

\bibitem{buttler} W.T.Buttler, R.J.Hughes, P.G.Kwiat, G.G.Luther,
 G.L.Morgan, J.E.Nordholt, C.G.Peterson, C.M.Simmons, Phys. Rev. A \textbf{57}, 2379 (1998), quant-ph/9801006.

\bibitem{tittel} W.Tittel, J.Brendel, H.Zbinden, N.Gisin
,Phys.Rev.Lett. \textbf{81} (1998) 3563-3566, quant-ph/9806043.

\bibitem{aspelmeyer}M. Aspelmeyer, T. Jennewein, M. Pfennigbauer, W. Leeb, A. Zeilinger,
IEEE Journal of Selected Topics in Quantum Electronics 1541- 1551,
quant-ph/0305105.

\bibitem{waks} E. Waks, A. Zeevi and Y. Yamamoto, Phys. Rev. A, Phys. Rev. A, {\bf 65},
52310 (2002), P. G. Kwiat, et al, Phys.Rev. A \textbf{60} (1999)
773-776.

\bibitem{bbm92} C. H. Bennett, G. Brassard and N. D. Mermin, Phys.
Rev. Lett. {\bf 68}, 557 (1992).

\bibitem{depol}A. Dragan, K. Wdkiewicz, Phys. Rev. A \textbf{71}, 012322 (2005), quant-ph/0407166.

\bibitem{Nielsen} M.A. Nielsen and I.L.Chuang Quantum Computation
and Quantum Information ( Cambridge University Press, Cambridge,
England, 2000).

\bibitem{preskill} Lecture notes on QC, J. Preskill,http://www.theory.caltech.edu/people/preskill/ph229.

\bibitem{wootters} W.K. Wootters, Phys. Rev. Lett. \textbf{80},2245
(1998).

\bibitem{horo} M. Horodecki, P. Horodecki, R. Horodecki, Phys. Rev. A \textbf{60}, 1888
(1999), quant-ph/9807091

\bibitem{classictele} F. Versraete, H. Verschelde, Phys. Rev. Lett. \textbf{90}, 097901
(2003), quant-ph/0303007

\bibitem{hor} P. Horodecki et al. Phys. Letts. A. {\bf 200} 340 (1995)

\end{thebibliography}
\end{document}